\documentclass[sn-mathphys-ay]{sn-jnl}

\usepackage{graphicx}%
\usepackage{multirow}%
\usepackage{amsmath,amssymb,amsfonts}%
\usepackage{amsthm}%
\usepackage{mathrsfs}%
\usepackage[title]{appendix}%
\usepackage{xcolor}%
\usepackage{textcomp}%
\usepackage{manyfoot}%
\usepackage{booktabs}%
\usepackage{algorithm}%
\usepackage{algorithmicx}%
\usepackage{algpseudocode}%
\usepackage{listings}%
\usepackage{textcomp}
\usepackage{setspace}
\usepackage{wrapfig}
\usepackage{subfigure}
\usepackage{wasysym}
\usepackage{derivative}

\theoremstyle{thmstyleone}%
\theoremstyle{thmstyletwo}%

\theoremstyle{thmstylethree}%

\begin{document}
\title[Article Title]{Active flow control over a sphere using a smart morphable surface}


\author[1]{\fnm{Rodrigo} \sur{Vilumbrales-Garcia}}\email{rodrigga@umich.edu}
\author[2]{\fnm{Putu Brahmanda} \sur{Sudarsana}}\email{brahmsdr@umich.edu}
\author*[1,2]{\fnm{Anchal} \sur{Sareen}}\email{asareen@umich.edu}

\affil[1]{\orgdiv{Department of Naval Architecture and Marine Engineering}, \orgname{University of Michigan, Ann Arbor}, \postcode{48109}, \state{MI}, \country{USA}}

\affil[2]{\orgdiv{Department of Mechanical Engineering}, \orgname{University of Michigan, Ann Arbor}, \postcode{48109}, \state{MI}, \country{USA}}


\abstract{Dimples on a sphere's surface can lead to significant drag reduction. However, the optimal dimple depth to minimize the drag varies with the Reynolds number ($Re = Ud/\nu$, where $U$ is the free stream velocity, $d$ is sphere diameter and $\nu$ is the fluid kinematic viscosity). To minimize drag across a wide range of $Re$ values, an active surface morphing approach is needed to adaptively adjust the dimple depth as incoming flow velocity changes. In this study, a smart surface-morphing technique is devised that can adjust dimple depth based on the flow conditions. By depressurizing the core of a rigid skeleton enclosed with a thin latex membrane, the dimple depth can be precisely controlled in response to flow velocity changes. A comprehensive series of systematic experiments are performed for Reynolds number range of $6\times10^4 \leq Re \leq 1.3\times10^5$, and dimple depth ratios of $0 \leq k/d \leq 2\times10^{-2}$ using a smart morphable sphere. It is observed that the dimple depth ratio $k/d$ significantly affects both the onset of the drag crisis and the minimum achievable drag. As $k/d$ increases, the critical Reynolds number for the drag crisis decreases. However, the minimum achievable drag coefficient decreases as $k/d$ increases. By carefully adjusting the $k/d$ to $Re$ using the morphable approach, our experiments show that $C_D$ reductions up to 50\% can be achieved when compared to a smooth counterpart for all the $Re$ considered in the study. For a constant $Re$, drag reduces as $k/d$ increases. However, there is a critical threshold beyond which drag amplification starts to occur. Particle image velocimetry (PIV) reveals a delay in flow separation on the sphere's surface with increasing $k/d$, causing the separation angle to shift downstream. This results in a smaller wake size and reduced drag. However, when $k/d$ exceeds the critical threshold, flow separation moves upstream, causing an increase in drag. This behavior closely resembles what has been observed in a smooth sphere experiencing drag crisis, where flow separation shifts upstream in the post-critical regime, leading to increased drag.
By using the experimental data, a control model is also developed relating optimum $k/d$ with $Re$ to minimize drag. This model also serves as the basis for adaptive drag control of the sphere for a wide range of Reynolds number. 

} 





\maketitle

\section{Introduction}\label{sec1}

Active surface morphing is ever present in nature. Animals can maximize their performance by carefully modifying their body to the surrounding environment, such as puffer fish and their body expansion when threatened by a predator \citep{wainwright1997evolution} and swift birds' wing morphing for maneuvering reasons \citep{lentink2007swifts}. The texture modifications of animals, whether passive or active, have served as inspiration for engineering applications for decades. For example, \cite{domel2018shark} found that passive rough elements inspired by shark-skin textures can reduce the drag generation of a wing, while \cite{van2015feather} reported that an aerofoil with roughness inspired by swifts can reduce flow separation. Another area of interest comprehending texture modifications for performance benefits involves the canonical problem of drag reduction of bluff bodies \citep{Bearman1976}. A bluff body experiences large drag due to the flow separation as a consequence of steep adverse pressure gradient. This can significantly affect the stability of structural loading which translates into aerodynamics efficiency of the body. One of the widely studied bluff bodies is sphere. A hallmark study which attempted to uncover the flow physics around sphere was conducted by  \cite{Achenbach1972} where a sphere was tested over a wide range of Reynolds number $5\times10^4 \leq Re \leq 6\times10^6$. A sudden decrease in the drag coefficient was observed at a critical Reynolds number of $3.7 \times10^5$. At critical Reynolds number, the boundary layer over the sphere surface transitions from laminar to turbulent delaying the flow separation leading to significant drag reduction, also known as the drag crisis. In the post critical regime, the flow separation location starts moving upstream that leads to an increase in drag \citep{Achenbach1972}.\\

Several studies have focused on investigating the effects of surface modifications on a sphere to reduce drag. Most techniques implement the concept of tripping the boundary layer by changing surface topology which induce flow instability and increase near-wall flow momentum that leads to the delay in flow separation \citep{Achenbach1974a,Choi2008,David2020, sareen2024}. Surface modification by dimples is one of the passive methods able to delay flow separation on a sphere \citep{Smits2004}. A systematic analysis of the drag performance on a dimpled sphere was firstly done by \cite{Davies1949} who studied the dynamics of a spinning golf ball, finding that adding dimples increased the flying distance of the ball. \cite{Bearman1976} reported that the addition of dimples reduced the critical $Re$ number when compared to a smooth counterpart, and performed a first analysis on the optimal shape of the elements, concluding that hexagonal dimples developed slightly lower drag than circular elements. The mechanisms behind the performance improvements was noted as the effectiveness of dimples in tripping the boundary layers and delaying the flow separation over the sphere surface. \cite{Choi2006} reported that the velocity fluctuations increase along the separating shear layer due to instability in the dimples, producing vortical structures and high momentum near the wall which overcomes the adverse pressure gradient effect and hence delays the flow separation. Similar flow behaviour has also been reported in a dimpled flat plate using computational approach \citep{Beratlis2014}.  A Direct Numerical Simulation (DNS) conducted by \cite{Beratlis2019} found that the mean turbulent kinetic energy starts to increase at position of $51^{\circ}$ due to dimples, leading to modifications in the velocity profiles and overcoming the effect of adverse pressure gradient. Nevertheless, the authors reported that, in the post-critical regime, dimples would decrease the mean skin friction coefficient and incur a local pressure penalty in the post-critical regime which increased the total drag force at about 50\%.\\

The general shape, size, and distribution of dimples over the sphere surface can have a large impact on drag reduction performance. For example, \cite{Aoki2003, AOKI2012} found that modifying the dimple depth can shift the critical Reynolds number $Re_c$ at which the drag crisis occurs. While deeper dimples were more effective in lowering drag at lower Reynolds numbers, but they increased drag at higher Reynolds numbers when compared to shallower dimples. 
Although the drag reduction capabilities of dimples on the surface of a sphere are well known, we can observe that those benefits are strongly correlated with the surrounding flow conditions. While previous studies have proven that significant drag reduction can be achieved, those are usually limited to a small range of $Re$, and a change in the flow conditions could quickly reduce any performance gains. To overcome this limitation, an active surface morphing strategy is needed to actively change dimple depth on-demand or adaptively with changing flow condition to minimize drag across a wide range of Reynolds numbers.\\

 \cite{Li2011} provided a theoretical basis of surface morphing and/or wrinkling transition of a core-shell soft sphere, which sufficiently matches with the drying mechanism of green peas. A control strategy for surface wrinkling was introduced by \cite{Breid2013} via UV exposure along with modification of the substrate geometric parameters. \cite{Terwagne2014} used the wrinkling instability of thin stiff films on curved compliant substrates through pneumatic actuation as an approach to generate intricate hierarchical surface patterns resembling labyrinth and hexagonal dimples and studied their effects on the drag coefficient. Their results suggest that the increase of roughness parameter of the patterns resembling hexagonal patterns shifts the occurrence of drag crisis to lower $Re$. Another surface morphing technique using pneumatic system was developed by \cite{Guttag2019} on a cylinder with dimples of various depths, finding a similar behavior in terms of drag crisis shifting as observed in a sphere with roughness \citep{Achenbach1974a} and with a dynamically changing trip wire \citep{Chae2019}. The use of grooves on the surface of a cylinder also reported the need of deeper elements at lower $Re$, in line with the previous studies \citep{Guttag2017}. Prior investigations into surface morphing techniques have supported the conclusion that roughness parameter (or dimple depth ratio in case of dimples) is an important factor in controlling the flow over a bluff body. \\

Previous research suggests that the ideal depth of dimples for drag reduction varies depending on the Reynolds number. Therefore, to achieve optimal drag reduction across a broad range of Reynolds numbers, it is necessary to develop a method that can adjust dimple depth adaptively as the Reynolds number changes. While \cite{Terwagne2014} proposed a surface morphing technique for altering surface topography on a sphere in response to varying flow velocities, this approach is complex to implement due to its reliance on a sophisticated wrinkling instability of thin stiff films on curved compliant substrates, activated by pneumatic means. This instability yields various surface patterns based on film stiffness, thickness and pressure actuation. Most of these patterns do not offer proven hydrodynamic benefits except for the hexagonal pattern resembling dimples.\\

In this study, we overcome these challenges and introduce a straightforward and simple surface morphing approach for drag reduction on a sphere, enabling precise control of dimple depth according to flow velocity. Utilizing this morphable sphere, we systematically investigate the impact of dimple depth ratio ($k/d$) on aerodynamic drag performance and flow characteristics. Although, several prior studies have studies the effect of dimples on the drag of a sphere and physical phenomenon associated with it \citep{Choi2006, Choi2008}, there is a lack of systematic investigation of the effect of dimple depth ratio on the aerodynamic drag and near-wake of the sphere. It is still unknown as to what are the optimal dimple depth ratios for varying Reynolds number? Furthermore, it is not known if for a fixed Reynolds number, increasing dimple depth ratio always increase drag or there is an optimal ratio beyond which the drag starts increasing? How the dimple depth ratio is correlated with the flow separation and near-wake of the sphere? This study addresses these unknown research questions and further advance our understanding of drag reduction using change in surface topography and its correlation with the near-wake and flow separation over the surface. \\

The present paper is organized as follows:  Section~\ref{sec2} describes the experimental methodology comprising experimental setup and development of the morphable sphere. Section~\ref{sec3} discusses the main findings of the experiment, while section~\ref{sec4} summarizes all the main findings of the present study and draw conclusions.

\section{Experimental Methods}\label{sec2}

The experiments were conducted in an open-loop wind tunnel at the Department of Aerospace Engineering at the University of Michigan, Ann Arbor with a test section of 0.6 m width, 0.6 m depth, and 3 m length. The wind tunnel was capable of producing steady flow speeds of $5 \leq U[m/s] \leq 22$. The free-stream turbulence intensity levels were within $T_i=1.8\%$ for the range of flow speeds tested. Although this value is relatively high compared to other studies, previous work has shown that $T_i<2\%$ have a limited impact on the drag production of a sphere in the sub-critical $Re$ range \citep{Achenbach1974a}. \\

\begin{figure}
    \centering
    \includegraphics[width=1\textwidth]{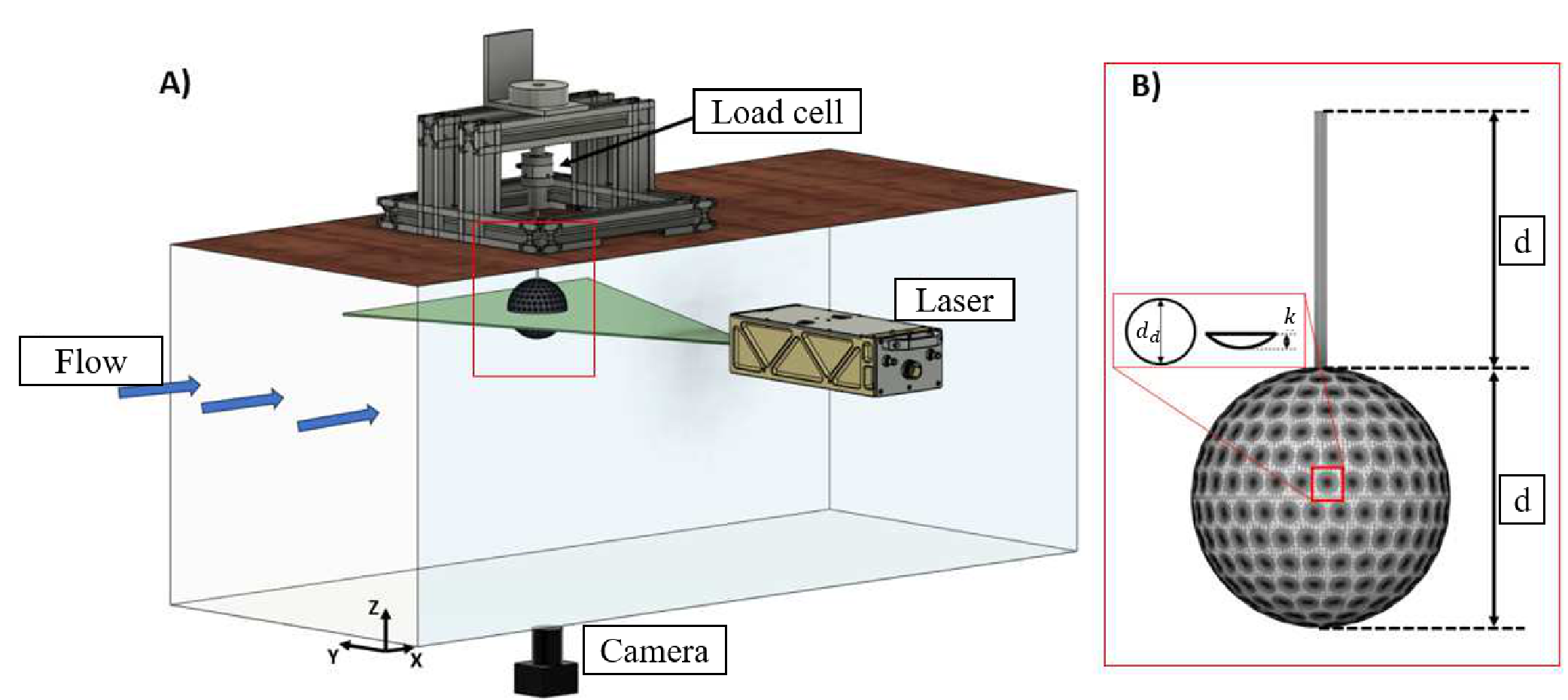}
    \caption{(A) A brief schematic (not to scale) of the experimental setup. (B) Schematic of the dimpled sphere model. Here, $d_d$ is the dimple diameter, $k$ is the dimple depth and $d$ is the sphere diameter.}
    \label{fig:fig1}
\end{figure}

A brief schematic of the experimental setup is shown in Figure~\ref{fig:fig1}. The sphere is mounted from the top of the test section with a holding structure containing a load cell. The sphere is supported with a cylindrical support rod of diameter $5$ mm, which is 20 times smaller than the sphere diameter. The exposed support rod length in the test section is equal to a sphere diameter $d$, in line with previous studies \citep{sareen2018vortex}. The blockage ratio, defined as the diameter of the sphere against the width of the test section ($d/W = 0.16$), is small enough to neglect the blockage effect on the flow field \citep{Achenbach1974a}.\\

The spheres considered in this study are constructed using a stereolithography (SLA) resin 3D printer (Form 3, FormLabs) with a resolution of $25$ $\mu$m. The morphology of the dimples are defined by two aspects: the dimple diameter ($d_d$) and the dimple depth ($k$). In line with \cite{Choi2006}, the ratio between the diameter of the dimple and sphere is kept constant at $d_d/d = 0.087$, while the normalized dimple depth ($k/d$) is varied from $0$ (smooth configuration), to $0.02$ in small increments. Two types of models are considered in this study: a sphere with rigid dimples and a morphable sphere with active dimples.
The former is built with similar non-dimensional parameters as \cite{Choi2006} for validation purposes ($d = 0.08$ m, $d_d/d = 0.087$, $k/d = 0.004$, and $384$ dimples uniformly distributed over the sphere surface ($N_d$). The sphere with morphing capabilities ($d = 0.1$ m, $d_d/d = 0.087$, $k/d = 0-0.02$, and $296$ dimples across the surface ($N_d$)) is built that can actively change its dimple depth on-demand. In the morphable sphere, the total number of dimples $N_d$ is reduced compared to \cite{Choi2006} to ensure structural integrity, leading to a coverage ratio (area of dimples against total surface area of the sphere) of  $\phi$ = $55.6\%$.\\

\begin{figure}
    \centering
    \includegraphics[width=0.9\textwidth]{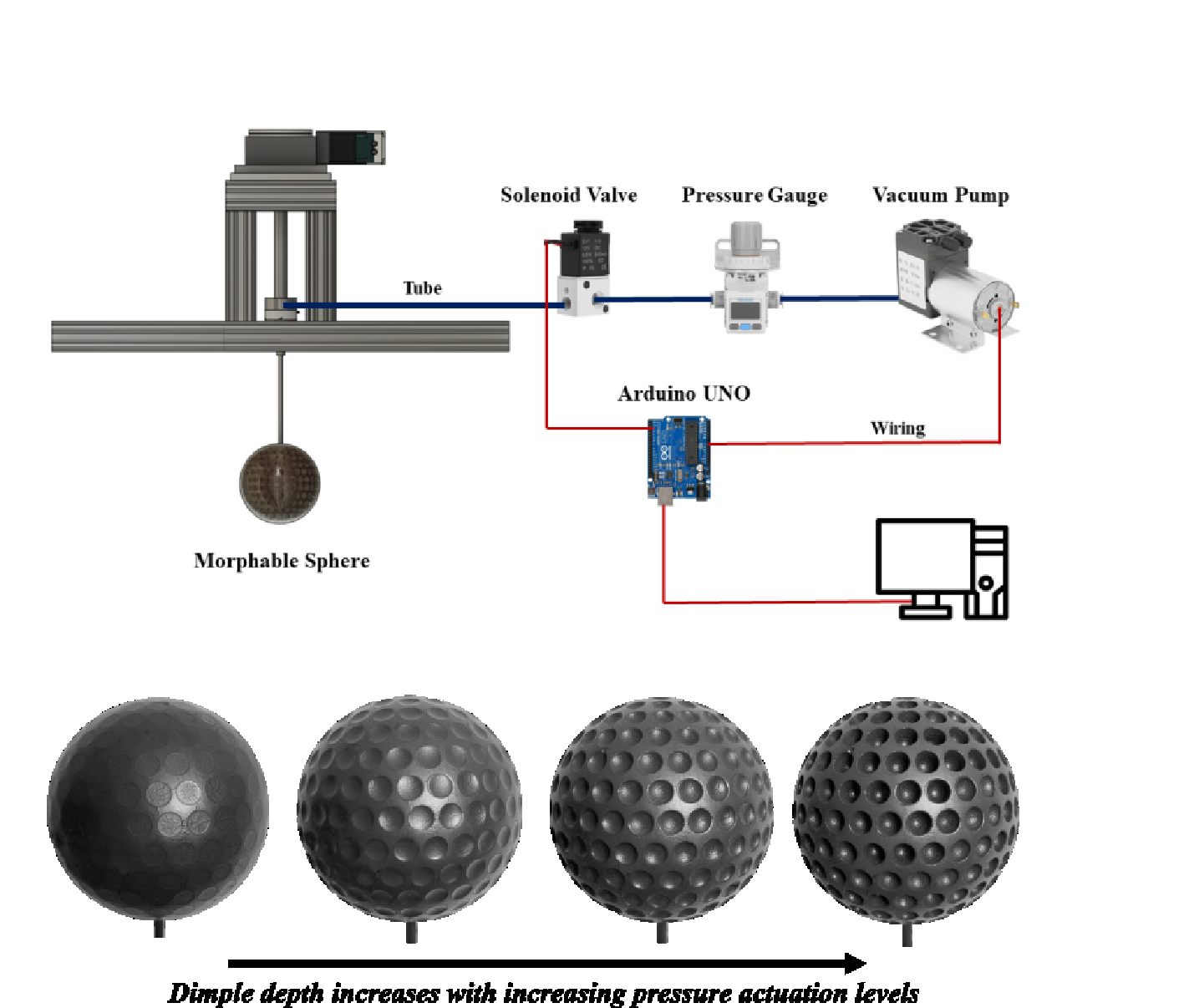}
    \caption{(Top) A schematic of the morphable sphere's pneumatic actuation system, which can actuate precise dimple depth on demand. (Bottom): An image of the morphable sphere shifting from a smooth (left) to a dimpled configuration with increasing dimple depth as depressurization levels in the core are increased utilizing the pneumatic actuation system. }
    \label{fig:fig2}
\end{figure}

\begin{figure}
    \centering
    \includegraphics[width=0.6\textwidth]{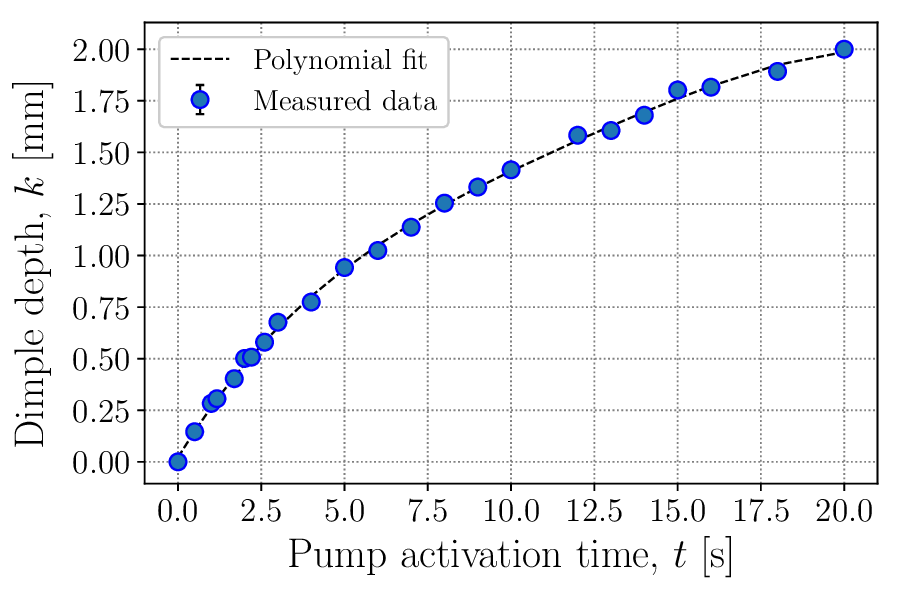}
    \caption{Calibration chart showing the variation of dimple depth $k$ [mm] as a function of the pump activation time [s]. The error bars represents the uncertainty in the measurements, and are mostly contained inside the markers. }
    \label{fig:fig3}
\end{figure}

The method for actuating dimples on the sphere's surface is outlined in Figure \ref{fig:fig2}. The morphable sphere consists of a 3D-printed rigid inner skeleton with circular holes uniformly distributed over its surface. The inner skeleton is covered by a pre-stretched thin latex membrane, as depicted in Figure \ref{fig:fig2}(a). Pre-stretching ensured that no fluttering of the latex membrane occurred during the testing. Dimples are actuated by depressurizing the sphere's core using a vacuum pump (Kamoer KLVP6) with a maximum pressure of $-85kPa$, controlled by an Arduino UNO. The dimple depth can be precisely controlled with a resolution of $0.01mm$ by varying the pump's activation time, illustrated in Figure \ref{fig:fig2}(b). Once the desired depth is attained, an Arduino-controlled solenoid valve seals the vacuum at the core of the sphere, maintaining the desired dimple depth. To establish the relationship between pump activation time and dimple depth ratio ($k/d$), the surface deformations are measured using a 3D scanner with a $5\mu m$ resolution. Dimple depth is recorded for various pump activation times, repeating the process three times for each dimple depth. Subsequently, as shown in Figure \ref{fig:fig3}, a polynomial fit is applied to the data points, yielding a control model to determine the necessary depressurization characteristics for achieving the desired $k/d$. It is important to note here that the maximum pressure due to the flow at the highest Reynolds number tested was only approximately $0.3 kPa$ which is an order of magnitude smaller than the vacuum pressure levels required to actuate the dimples. Thus, any surface deformation over the sphere surface due to the flow can be neglected. \\

The forces acting on the spheres were measured with a six-axis force sensor (ATI mini40 IP65), with a resolution of $0.01$ N in the streamwise and spanwise force direction. The forces were acquired at a frequency of $1 kHz$ , for a duration of 1 minute, with each data point repeated at least three times. The non-dimensional time, defined as $t^*=tU_\infty/d$ varied from 5300 to 11500 across the $Re$ range considered. As shown by \cite{Norman2011}, a $t^*>2000$ is required to achieve converged statistics for a sphere in the sub-critical $Re$ range. In line with the previous findings, our $t^*$ values are at least $2.5$ times higher, ensuring converged forces. The reported streamwise force (drag) is defined as,

\begin{equation}
    C_D=\frac{2 F_D}{\rho U_\infty ^2 A}
\end{equation}

Where $F_D$ represents the drag force in N, $\rho$ is the fluid density, $U_\infty$ denotes the free-stream flow velocity, and $A$ is the projected area of the sphere, calculated as $A=\pi d^2/4$. The data acquisition and analysis was performed using Matlab and Python. The effects of the supporting rod have been isolated from the sphere by linearly subtracting its drag production assuming a constant $C_D=1.2$ for the $Re$ range considered \citep{SAREEN2018, sareen2024}. The uncertainty in the measurements $u$ has been calculated following the procedure stated by \cite{taylor1982introduction}:

\begin{equation}
    u=\sqrt{\left(\pdv{C_D}{F_D} u_f\right)^2 + \left(\pdv{C_D}{U} u_u\right)^2}
\end{equation}

where $u_f$ and $u_u$ correspond to the uncertainty in the forces and velocity readings. The values obtained with the previous equation are later combined with the standard error ($SE=std(C_D)/\sqrt{N}$), where $N$ is the amount of repetitions, as follows: $u_{C_D}=\sqrt{u^2+SE^2}$
\\

The flow velocity fields were captured using 2D-2C Particle Image Velocimetry (PIV). The laser plane is aligned along the streamwise direction ($XY$ plane), passing through the centre of the sphere. The flow was seeded with polydisperse aerosol by atomizing Di-Ethyl-Hexyl-Sebacat (DEHS) solution into particles
using LaVision Aerosol Generator. DEHS particles have a mean size of $1 \mu m $ and a density of $0.91 g/cm^3$. The particles were illuminated with an Evergreen 200 dual-pulsed laser with a pulse energy of $200 mJ$ and frame rate of $15 Hz$. The PIV images were captured with a LaVision Imager CX-5 camera with a resolution of 2440 $\times$ 2040 pixels$^2$. The camera was equipped with a $60mm$ Nikon lens to give a field of view of $113 \times 95 mm $. DaVis 11 software is used to cross-correlate the acquired particle image pairs with an interrogation window size of $64$x$64$ pixels with a 50\% overlap, providing a spatial resolution of $0.03d$. The images were captured for one minute at a rate of $15Hz$, totaling 900 image pairs, to achieve statistically robust results.\\

The matrix of characteristic parameters followed in this study are presented in Table \ref{tab:cases}. Experiments were conducted for a Reynolds number varying in the range of $Re= DU_\infty/\nu$=$60, 000$- $130, 000$, where $U_\infty$ is the free-stream velocity, $\nu$ is the kinematic viscosity, and $d$ is the sphere diameter. Two types of spheres are considered in this study. The first one consists of a sphere with rigid dimples for validation purposes ($d=0.08 m$). The second one consists of a morphable sphere of diameter $d=0.1m$. The normalised diameter of the dimple ($d_d/d$) and the area coverage ratio $\phi$ = $55.6\%$ is kept constant across all the cases considered. The normalised dimple depth ($k/d$) is varied in fine resolutions from 0 (smooth configuration), to $0.02$ (dimple depth = $2 mm$).

\begin{table*}
    \centering
    \begin{tabular*}{0.75\textwidth}{@{\extracolsep{\fill}} l l l}
    \hline \\
         Reynolds number & $Re = U d/\nu$ & 60, 000 - 130, 000 \\ [0.25cm]
         Dimple depth ratio & $k/d$ & 0.000 - 0.02 \\[0.25cm]
         Sphere diameter & $d$ & 0.08 , 0.1 m \\[0.25cm]
         Dimple diameter & $d_d$ & 8.67 mm \\[0.25cm]
         Dimple depth & $k$ & 0 - 2 mm \\[0.25cm]
         Area coverage ratio & $\phi = N_d  d_d^2/(4d^2) $ & 55.6\%\\[0.25cm]
        \hline\\
         
    \end{tabular*}
    \caption{Table showing matrix of characteristic parameters followed in the present study. Here, $U$ is the flow velocity, $d$ is sphere diameter, $\nu$ is kinematic viscosity of air, $k$ is the dimple depth, $d_d$ is the dimple diameter and $N_d$ is the total number of dimples.}
    \label{tab:cases}
\end{table*}

 \section{Results and Discussion}\label{sec3}

This section discusses the results obtained in this experimental study. First, we start by validating the experimental setup and the morphing strategy with previous studies using spheres with rigid dimples. Next, we investigate the effect of systematically varying the dimple depth ratio or roughness parameter ($k/d$) using the morphing strategy, and compare the evolution of drag coefficient ($C_D$) across a range of Reynolds numbers ($Re$) values. Finally, we discuss the forces results and correlate them with Particle Image Velocimetry measurements.

\subsection{Validation}
\begin{figure}
    \centering
    \includegraphics[width=0.85\textwidth]{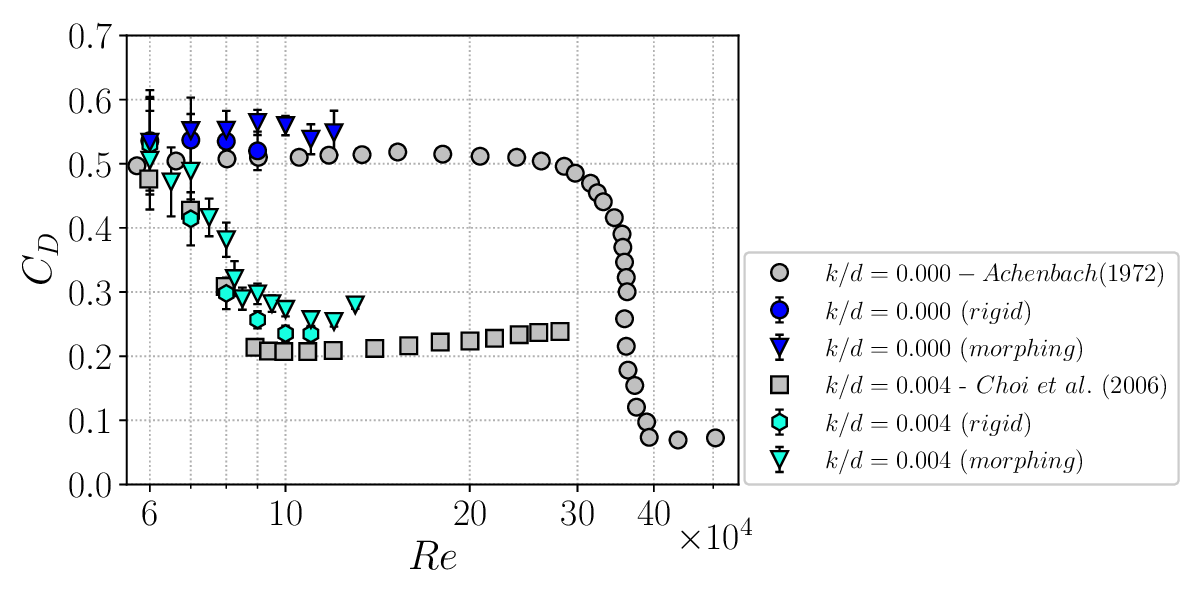}
    \caption{$C_D$ against $Re$ evolution of rigid and morphable sphere compared with previous studies \citep{Achenbach1972, Choi2006}.}
    \label{fig:fig5}
\end{figure}

For the validation, we compare the drag force on a smooth and rigid dimpled sphere with prior benchmark studies. Figure \ref{fig:fig5} depicts the evolution of $C_D$ for various $Re$ values, comparing a smooth sphere ($k/d=0.000$) with a sphere featuring fixed dimples at $k/d=0.004$. Our findings are juxtaposed with previous studies with similar characteristics \citep{Achenbach1972, Choi2006}. For the smooth configuration, we observe a constant $C_D\approx 0.5$ for $Re<100,000$, aligning with prior seminal research work of \cite{Achenbach1972}. Introducing dimples triggers laminar to turbulent boundary layer transition leading to delay in the flow separation over the sphere surface \citep{Choi2006}. This leads to an earlier onset of drag crisis with a notable reduction in $C_D$ for $Re$ values after the drag crisis. This effect is evident in the drag characteristics of the sphere with fixed $k/d=0.004$ dimples. Consistent with \cite{Choi2006}, we observe a $C_D$ crisis starting at $Re=70, 000$, and reaching a minimum $C_D$ of 0.23 at $Re=100,000$. While our minimum $C_D$ value is slightly higher than that reported by \cite{Choi2006}, our results exhibit a comparable trend.\\

\begin{figure}
    \centering
    \includegraphics[width=0.6\textwidth]{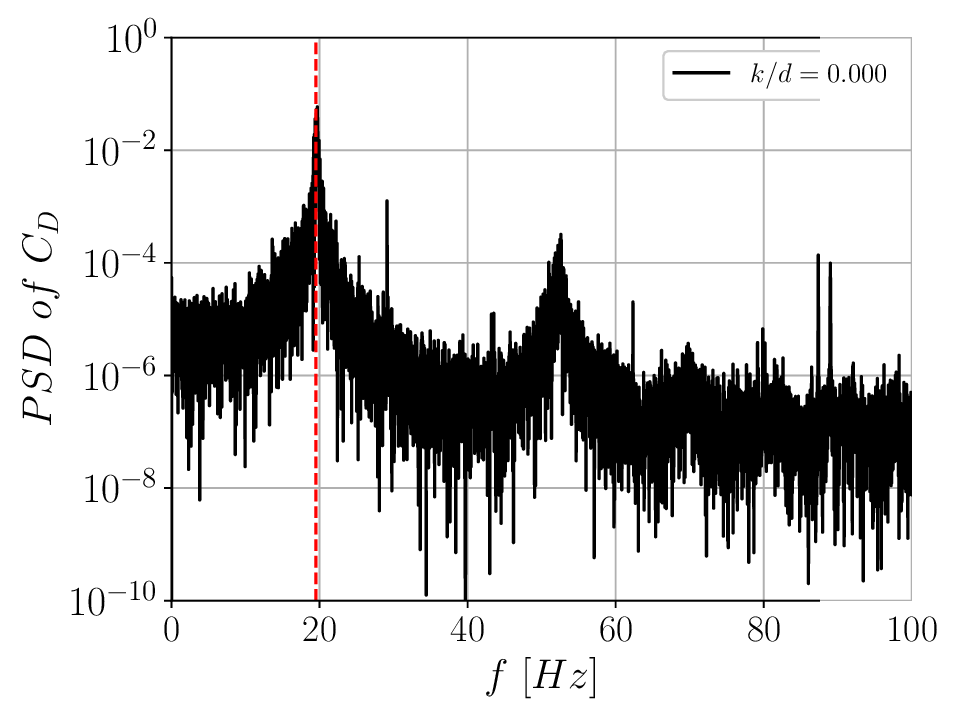}
    \caption{PSD analysis of a morphable sphere under smooth configuration at $Re=60, 000$. The dashed red line denotes the dominant frequency in the drag force signal.}
    \label{fig:fig4}
\end{figure}

To ensure that the morphing strategy can generate and maintain a fixed dimple depth that compares well with its rigid counterpart for the Reynolds numbers tested in the study, we also compare drag force on a dimpled sphere with rigid dimples to a dimpled sphere actuated using the morphable approach.
In Figure~\ref{fig:fig4}, we compare the $C_D$ values between a sphere with rigid dimples and fixed dimples obtained via the morphable approach at $k/d=0.004$. We find good agreement, particularly at $Re>100,000$, both in overall drag values and in the trend of the drag crisis. However, at lower $Re$, the morphable sphere exhibits a slightly higher $C_D$, possibly attributed to differences in the number of dimples present on the surface, since the area coverage ratio, or the total area of the surface covered by rough elements, has been previously noted to influence the sphere's $C_D$ performance \citep{Aoki2003}. Nonetheless, the similarities in the drag crisis trend validate the morphing approach, confirming the morphing system's ability to achieve the expected $k/d$ with high precision.\\
 
Figure \ref{fig:fig4} shows a Power Spectra Density ($PSD$) on the raw $C_D$ signals for a morphable sphere with smooth configuration at $Re=60, 000$. As evident from the figure, there is a characteristic peak corresponding to the vortex shedding frequency, highlighted by a red dashed-line located at around $19Hz$, leading to a Strouhal number ($St=df/U_\infty$), where $d$ is the sphere diameter, and $f$ is the vortex shedding frequency, of $St=0.21$, in line with previous studies by \cite{Achenbach1974a}. \\

In summary, the results discussed in this section provides validation for the current experimental setup and the morphing strategy. In the following, we will discuss the effect of roughness parameter or the dimple depth ratio on the drag of a sphere for the Reynolds number range $Re = 60 000 - 130 000$. We will then correlate them with the PIV results. Finally, we will employ the experimental results to develop a model relating optimal $k/d$ and $Re$ for minimizing drag across the entire range of Reynolds numbers tested in the study. This control model will then be used to demonstrate adaptive morphing capabilities developed in the current study. 

\subsection{Drag of a morphing sphere for varying $k/d$ and $Re$}
\label{sec3.2}

\begin{figure}
    \centering
    \includegraphics[width=1\textwidth]{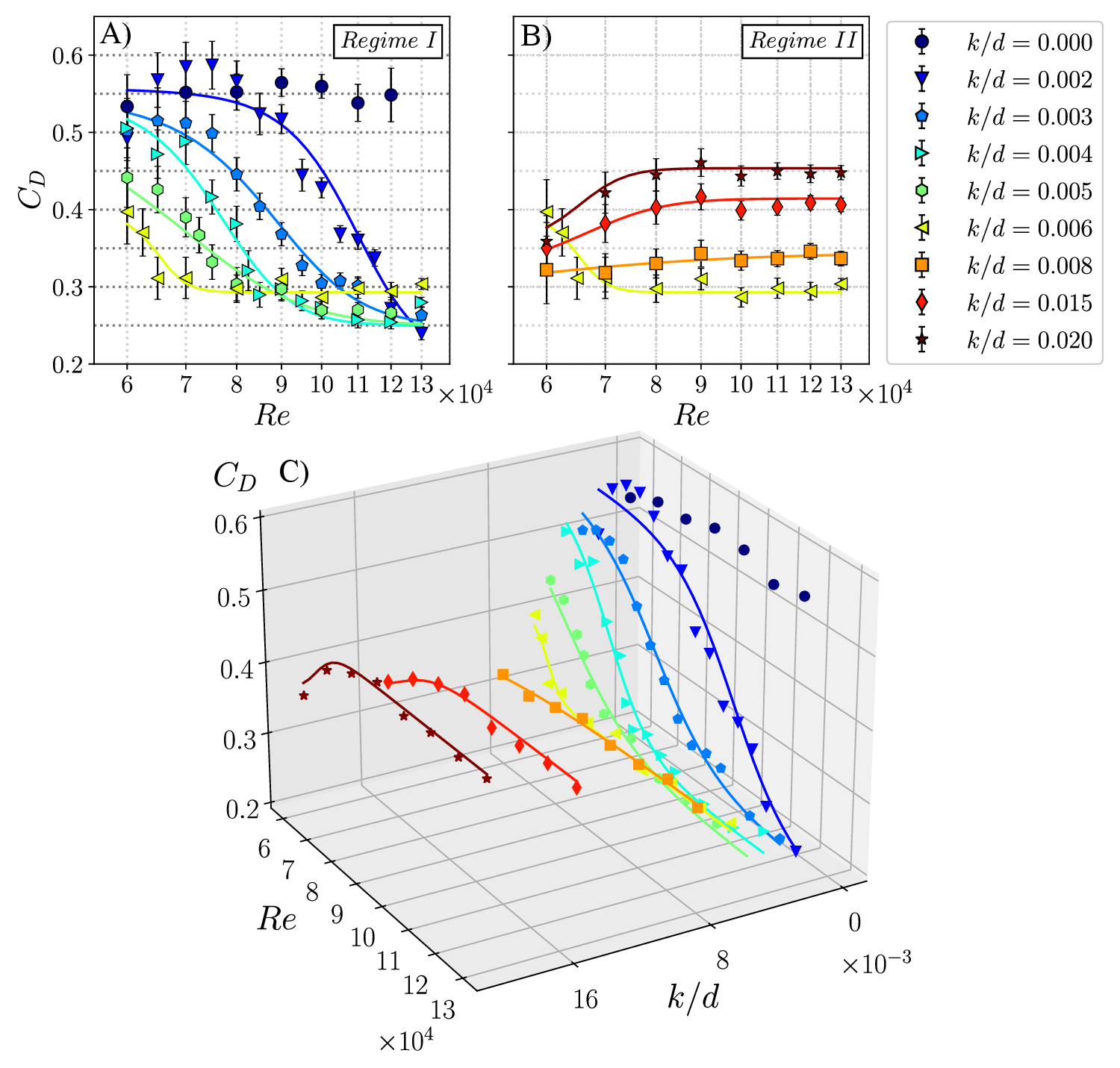}
    \caption{(Top) $C_D$ against $Re$ evolution for all the $k/d$ considered in this study. Figure on the bottom shows the results in an auxiliary 3D view. Blue colours indicate shallower dimples, and red colours note deeper dimples.} 
    \label{fig:fig6}
\end{figure}

Figure \ref{fig:fig6} presents the evolution of the drag coefficient ($C_D$) against the Reynolds number ($Re$) for dimple depth ratios  ($k/d$) varying from  0 (corresponding to a smooth configuration) to $k/d=0.02$. Overall, two discernible trends emerge with increasing $k/d$, as shown separately in Figure~\ref{fig:fig6}A and \ref{fig:fig6}B. In Regime I, as illustrated in Figure \ref{fig:fig6}A, as $k/d$ increases from $0.02\leq k/d\leq0.006$, the critical Reynolds number $Re_c$ decreases. 
The critical Reynolds number at which the drag rapidly drops reduces with increasing $k/d$, reaching values of $Re=70,000$ for $k/d=0.006$. This behavior has also been reported previously for other surface modifications, such as roughness and surface trips \citep{Choi2006, Chae2019}.\\

Following the attainment of lowest $C_D$, further increase in $Re$ at constant $k/d$ appears to exert limited influence, as the evolution of $C_D$ stabilizes across all the dimple depth ratios studied, which is presumably due to constant flow separation angle on the sphere surface in the post-critical $Re$ range, as also reported by \cite{Choi2006}. It is also noticeable how the minimum $C_D$ that can be achieved with the addition of dimples decreases as $k/d$ is increased in this regime. The sphere with shallowest surface modifications, $k/d=0.002$, leads to a minimum drag of $C_D=0.25$ at $Re = 130000$, resulting in 50\% drag reduction when compared with its smooth counterpart. \\

As evident from Figure \ref{fig:fig6}A, there is no one optimal $k/d$ that minimizes drag for the entire range of Reynolds numbers tested. For lower Reynolds number, deeper dimples are more effective in reducing drag, however, as $Re$ is increased, shallower dimple depth are required to minimize the drag. Thus, if our goal is to minimize drag for the entire range of Reynolds numbers, we need a strategy whereby we can adapt dimple depth with the flow condition. By doing so, it will be possible to reduce the $C_D$ of a sphere by at least 40\% across a $Re$ range spanning from $60,000$ to $130,000$.\\

In Regime II, shown in Figure~\ref{fig:fig6}B, when $k/d$ is further increased from $0.006<k/d\leq 0.02$, a reversal in this trend is observed, whereby the drag starts increasing. While for $k/d=0.006$, the minimum $C_D$ of 0.3 is observed at $Re=65,000$, higher $k/d$ preclude the drag crisis within the range of $Re$ considered. Overall, we observe that $C_D$ reductions decrease as $k/d$ is augmented more than $k/d=0.006$, with the trend in drag against $Re$ shifting up, back towards that of a smooth sphere. The deepest dimple depth analysed in this study, $k/d=0.02$ reports $C_D$ values around $0.46$, which constitutes a mere 9\% drag reduction when compared to the smooth configuration. We also note that the effects described at Figure \ref{fig:fig6}B highlight the importance of controlling $k/d$ under specific flow conditions, as adjustments in this parameter can swiftly negate any potential performance advantages.\\

\begin{figure}
    \centering
    \includegraphics[width=1\textwidth]{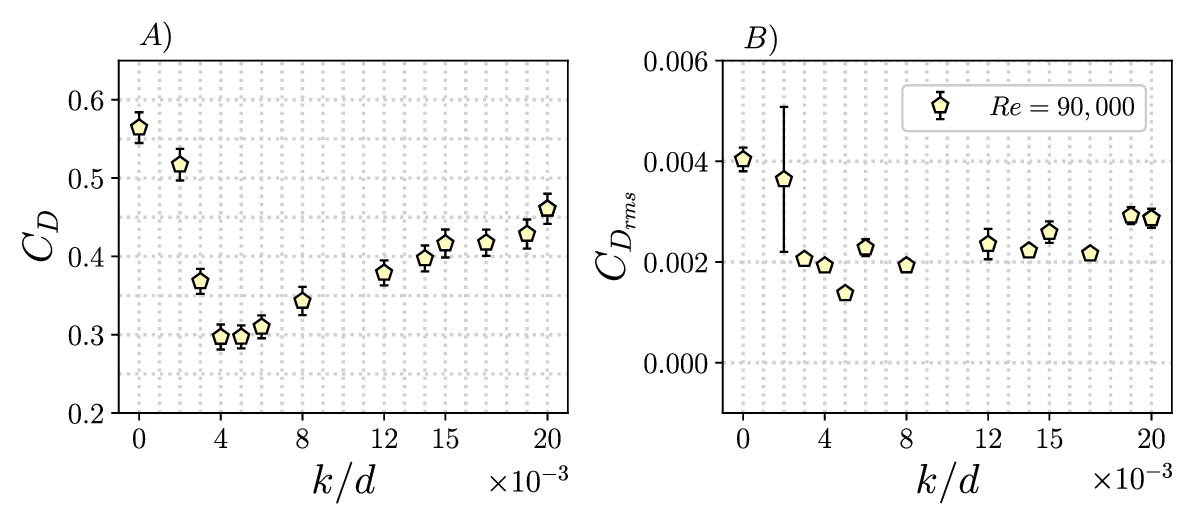}
    \caption{A) $C_D$ against $k/d$ evolution at $Re=90, 000$. B) RMS  of the instantaneous $C_D$ at $Re=90, 000$.  }
    \label{fig:fig8}
\end{figure}

The effects explained in Figure \ref{fig:fig6} can be further highlighted by analysing the $C_D$ evolution against $k/d$ at fixed $Re$ number, as shown in Figure \ref{fig:fig8}. Here, we report the drag of the morphable sphere with varying $k/d$ at $Re= 90, 000$. As $k/d$ increases from 0 to 0.005, the drag rapidly drops reaching maximum reduction of up to almost 50\%. A further increase in $k/d$ beyond this optimal value leads to a rapid increase in drag eventually approaching that of a smooth sphere. This corroborates that there is an optimal $k/d$ for each Reynolds number that maximizes the drag reduction. Previous research (and our own observations discussed in the wake measurement section), has established that the drag crisis of a dimpled sphere occurs when the boundary layer transitions from laminar to turbulent. This transition delays the flow separation, thereby reducing drag. However, when the roughness parameter (in this case, the dimple depth ratio) exceeds a critical threshold, drag begins to increase, indicating that the location of flow separation moves upstream. This phenomenon is analogous to the behavior observed in smooth spheres in post-critical regime where the drag begins to increase as flow separation location shifts upstream. Our current research demonstrates a similar trend with increasing roughness parameters, representing a novel discovery that suggests the potential for manipulating wake deflection around a sphere solely through roughness. \\

We also report the root-mean-square values of the time series of the drag signal in Figure~\ref{fig:fig8}B. As evident from the figure, $C_{D rms}$ rapidly drops with increasing $k/d$ reaching almost 50\% lower $C_{Drms}$ compared to that of a smooth sphere. Interestingly, it seems that, after reaching the $k/d$ for minimum $C_{D rms}$, the values remain stable for higher $k/d$ although the time-averaged drag coefficient starts increasing. This finding points towards potential capabilities of dimples to not only reduce the drag of a sphere, but also to significantly limit force fluctuations, with $rms$ values of less than 40\% of those produced by a smooth configuration. 

\subsection{Near-wake measurements using Particle Image Velocimetry}
After describing and analysing the drag reduction capabilities of the morphable sphere actuated with various fixed dimple depths, we aim to further explain the reasons behind any performance augmentation by drawing relations between forces and wake characteristics measured with 2D-2C Particle Image Velocimtery (PIV). The PIV is performed in an equatorial plane passing through the center of the sphere. The details of the PIV setup are given in Section~\ref{sec2}.\\

\begin{figure}
    \centering
    \includegraphics[width=1\textwidth]{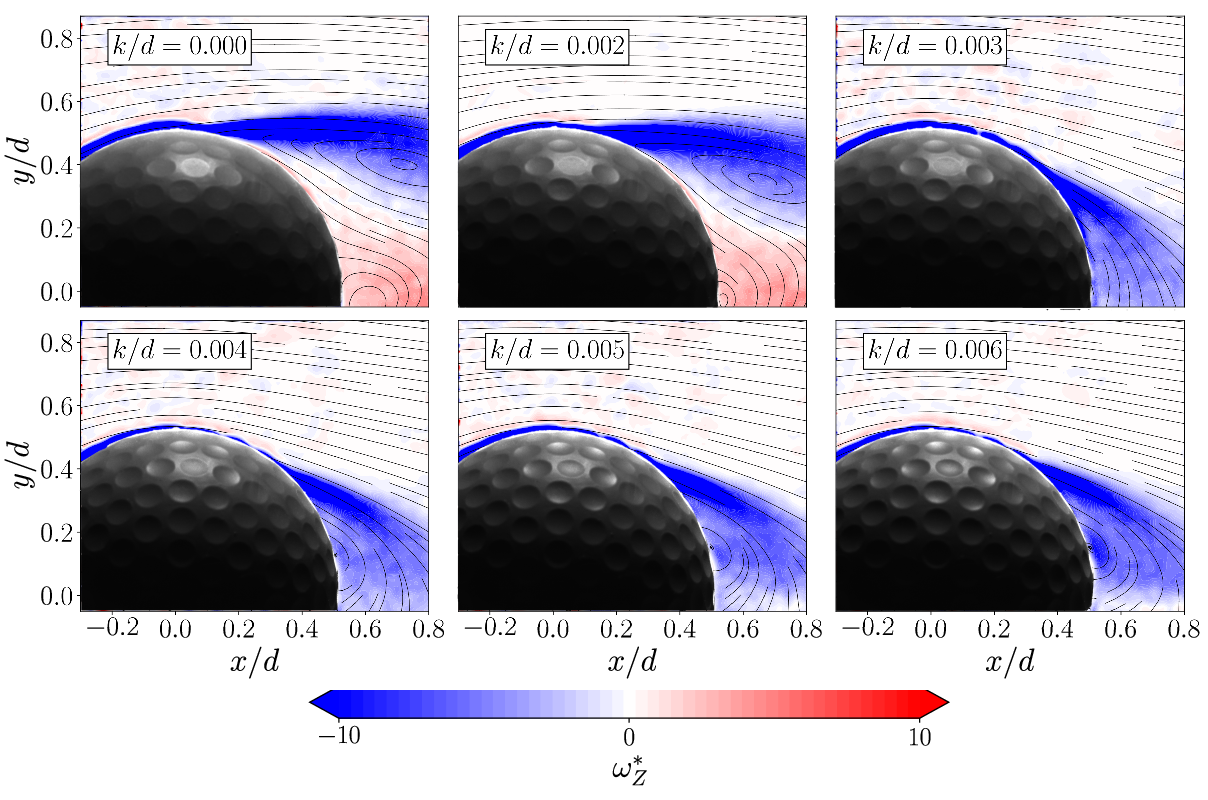}
    \caption{Normalised vorticity $\omega_z^*$ flow-fields and time-averaged streamlines for several $k/d$ at $Re=90, 000$. }
    \label{fig:fig9}
\end{figure}

Figure \ref{fig:fig9} displays the normalized vorticity fields $\omega_z^*$ overlaid with time-averaged streamlines for dimple depth ratios of $k/d=0-0.006$ at a Reynolds number of $Re=90,000$. As evident from the figure, when dimple depth ratio increases from $k/d = 0$ to $k/d=0.002$, the near-wake characteristics and the flow separation location remains largely unchanged. This is consistent with the force data shown in Figure~\ref{fig:fig6}. However, for $k/d = 0.003$, notable narrowing of the wake region is clearly evident associated with significant delay in the flow separation location. This is correlated with the substantial $C_D$ reduction (26\%) noted earlier.\\

Transitioning to $k/d=0.004$ yields a further $C_D$ reduction of 18\%, yet the wake topology appears similar to $k/d=0.003$. As reported by \cite{Beratlis2019}, the wake shape and separation point location may not necessarily correlate with $C_D$ reduction, and future computational work aimed at flow measurements very close to the dimples may be necessary to elucidate this aspect. This is particularly very challenging endeavor for experiments. Consistent with $C_D$ trends, $k/d=0.005$ and $k/d=0.006$ do not induce further reductions in wake size compared to $k/d=0.004$, and appears to slightly move the separation angle frontwards, in line with the trend reversal in the forces denoted before. \\

\begin{figure}
    \centering
    \includegraphics[width=0.8\textwidth]{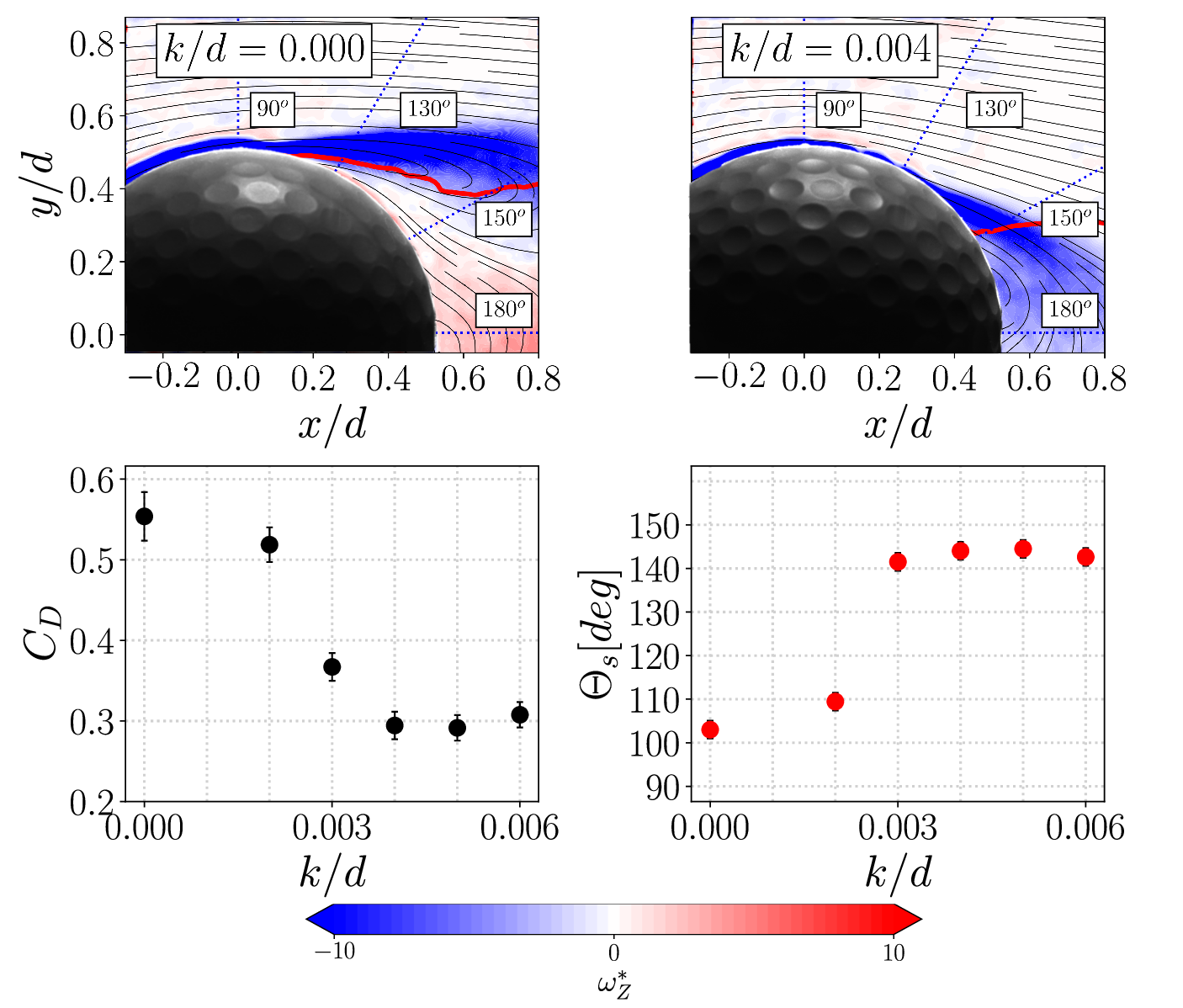}
    \caption{Normalised vorticity $\omega_z^*$ flow-fields and time-averaged streamlines for $k/d=0.000$ and $k/d=0.004$ at $Re=90, 000$ (top row). $C_D$ and $\theta_s$ for several $k/d$ (bottom row).}
    \label{fig:fig10}
\end{figure}

To further examine the wake characteristic for various $k/d$ ratios, we illustrate in Figure \ref{fig:fig10} a comparison between $C_D$ and the global flow separation angle $\theta_s$ (measured from the leading stagnation point) of the flow on the surface of the sphere. We determine $\theta_s$ by identifying the point on the body's surface where $U_\theta=0$, referencing it to the stagnation point on the sphere ($y/d=0$, $x/d=-0.5$) \citep{Moore1958,Kim2014}. The uncertainty in $\theta_s$ is obtained by calculating the angle given by the window size considered in the PIV processing. Figure \ref{fig:fig10} presents the averaged normalized vorticity $w_Z^*$ overlaid with time-averaged flow streamlines. A contour line where $\theta_s=0$ is also highlighted in red color. The resulting $\theta_s$ and $C_D$ values are also displayed at the bottom row of Figure \ref{fig:fig10}. As previously discussed, for a given Reynolds number ($Re$), there exists an optimal $k/d$ ratio for maximizing drag reduction. Deviating from this optimal ratio, either by decreasing or increasing $k/d$, rapidly diminishes performance benefits. At $Re=90,000$, $k/d=0.004$ yields the minimum $C_D=0.3$, corresponding to a separation angle of $\theta_s=147^\circ$. This reduction in $C_D$ compared to its smooth counterpart amounts to 40\%. Correspondingly, the separation angle is delayed by $50^\circ$. Although our measured $\theta_s$ differs from previous data (with $\theta_s=97^\circ$ for smooth surfaces and $\theta_s=147^\circ$ for $k/d=0.004$, while \cite{Choi2006} reported $\theta_s=82^\circ$ and $\theta_s=110^\circ$ for similar $k/d$), the strong correlation between $C_D$ and $\theta_s$ suggests that any disparities may stem from variations in sphere dimple arrangements \citep{Aoki2003}. Notably, our $k/d=0.004$ features a lower dimple coverage ratio compared to \cite{Choi2006}, while the $k/d=0.000$ configuration for the morphable sphere exhibits shallow dimples due to pre-stretching of the flexible membrane. As $k/d$ is further increased, $\theta_s$ begins to converge back towards the value observed in the smooth configuration, remarking the trend reversal found during the forces analysis.
Overall, Figure~\ref{fig:fig10} establishes that changes in drag are correlated with changes in the global flow separation location in the subcritical Reynolds number regime tested in this study. 

\subsection{Model relating optimal $k/d$ and $Re$ for minimizing drag}
\begin{figure}
    \centering
    \includegraphics[width=0.8\textwidth]{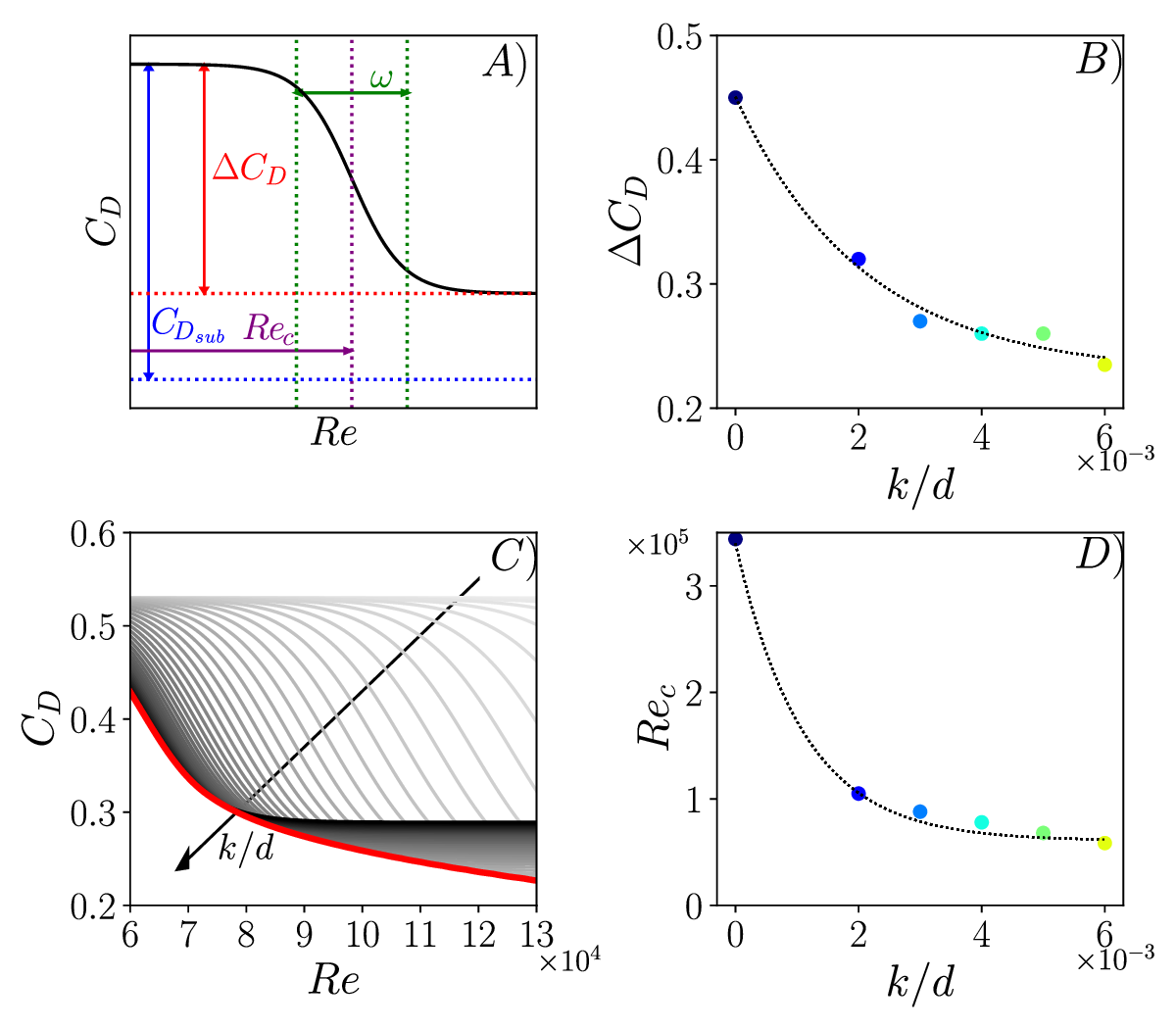}
    \caption{A) Schematics of the main parameters used to fit the $C_D$ vs $Re$ model for different $k/d$ (reproduced from \cite{Chae2019}). B) $\Delta C_D$ against $k/d$, and exponential fit used to obtain the curves at A) with Eq. \ref{Eq3}. C) $C_D$ against $Re$ estimation for several $k/d$. The red line denotes the optimal curve of minimum $C_D$.  D) $Re_c$ against $k/d$, and exponential fit used to obtain the curves at A) with  Eq. \ref{Eq3}.}
    \label{fig:fig7}
\end{figure}

In this section, we utilize the data presented in Section~\ref{sec3.2} to develop a control model relating optimal $k/d$ and $Re$ for minimizing drag. To do so, we follow the approach presented in \cite{Chae2019}. The authors provide an empirical model relating the $C_D$ evolution with $Re$ as follows:

\begin{equation}
    C_D(Re)=C_{D,sub}-\mid \Delta C_D \mid \left(1+exp\left[\frac{-2ln(1/s-1}{\omega}(Re-Re_c)\right]\right)^{-1}
    \label{Eq3}
\end{equation}

where $C_{D,sub}$ is the $C_D$ value in the subcritical Reynolds numbers before the drag crisis, $\omega$ is the change in  $Re$ between the pre and post-crisis, $Re_c$ is the critical Reynolds number at which drag crisis occurs, and $s$ denotes the steepness of the drag crisis curve, as shown in Figure~\ref{fig:fig7}. Due to limitations in the $Re$ range of this study, we utilise the same values as \cite{Chae2019} for the smooth curve. In line with \cite{Chae2019}, $s$ is kept constant as $0.1$, since it provides a good fitting for all the curves considered. The previous equation assumes the existence of a sub-critical and post-critical $Re$, which limits its applicability in our study to $k/d<0.006$. Nevertheless, $k/d>0.008$ are not providing any $C_D$ reduction for the $Re$ considered, making their consideration irrelevant for a drag reduction model. The data acquired during the experimental campaign is used to find the parameters that produce an optimum fit between $C_D$ and $Re$. In Figure \ref{fig:fig7}, we introduce the main parameters used to produce $C_D$ vs. $Re$ trends for all the spheres, with all the curves fitting the data with a $r^2$ value higher than $0.9$. Once the main parameters are found, we can produce relations between $\Delta C_D$, and $Re_c$ with $k/d$. By doing so, we can project the $C_D$ evolution against $Re$ for any $k/d$, leading to the curves presented at Figure \ref{fig:fig7}. Finally, we can use the new projections to delineate a relation between $k/d$ and $Re$ for minimum drag production, as depicted by a red line at  Figure \ref{fig:fig7}:

\begin{equation}
    k/d_{opt}(Re)=0.022 e^{(-2.05\times 10^{-5} Re)}+4.06\times 10^{-4}
    \label{Eq4}
\end{equation}

This model not only provides a useful relationship between optimal roughness parameter and Reynolds numbers for minimizing drag, but can also serve as the basis for a real-time closed-loop control implementation that is able to deploy optimum dimple depths adaptively with changing $Re$ conditions. In the following section, we utilize this model to showcase real-time close loop implementation, whereby the morphable sphere detects flow velocity and adjusts dimple depth automatically in real-time to minimize the drag. \\

\subsection{Closed-loop implementation of optimal dimple depth for minimizing drag}

As described before, the main advantage of our smart morphable approach relies on the fact that it can adjust the surface topography and tune it to the oncoming $Re$ to achieve performance gains. In this section, we present a first approach towards closed-loop control aimed at maximising performance across a wide range of incoming flow conditions. \\

The only parameter needed for a closed-loop control approach in this case is the incoming flow velocity, or $Re$. Once $Re$ is known, we can estimate the optimum dimple depth needed for drag reduction using the model shown in Equation~\ref{Eq4}. This usually necessitates adding an additional sensor to measure flow velocity that increases the system complexity and cost. To overcome this, we devised an alternative approach. The drag coefficient of a smooth sphere is constant ($C_D \approx 0.55$) in the subcritical regime. We use this information to get an estimate of flow velocity from the instantaneous force sensor data. Once $Re$ is estimated, we can deploy the dimples to the optimum $k/d$ using the pneumatic actuating system. The main limitation of this approach relies on the fact that, if the velocity changes, the sphere needs to return to the smooth configuration to get a new velocity estimation. Nevertheless, this approach serves as a good preliminary model to showcase the adaptive drag reduction capabilities of the morphing strategy developed in this study. An alternative model for flow velocity estimation would be to use a gradient based optimization technique to avoid reverting back to smooth configuration every time the flow velocity is changed. This will be the focus of the future follow-up study.  \\

\begin{figure}
    \centering
    \includegraphics[width=1\textwidth]{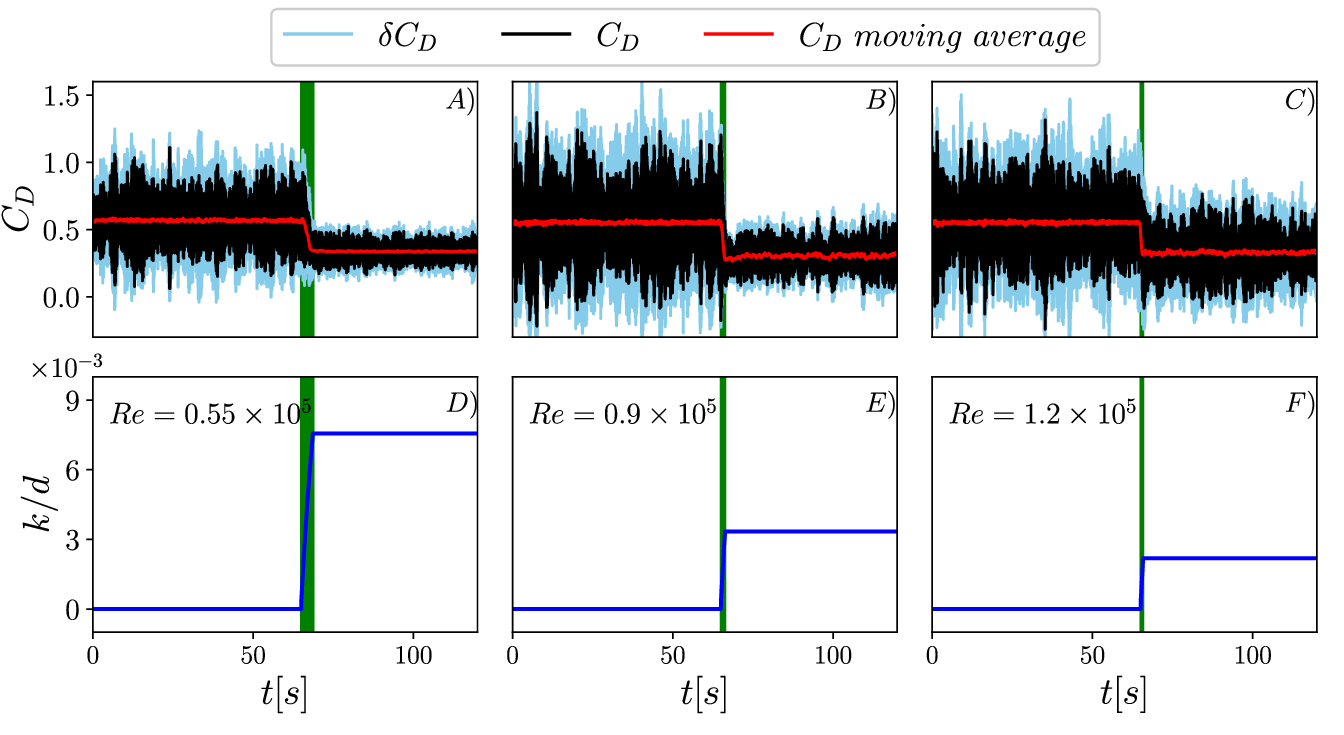}
    \caption{$C_D$ evolution in time demonstrating closed-loop implementation for drag reduction purposes. Black line indicates instantaneous $C_D$ averaged across three trials (A, B, C) and $k/d$ evolution as prescribed by the controller (D, E, F). Blue regions represent standard deviation across the three trials. Red line presents a moving average with a window of $1$ second. A) and D) for $Re=55, 000$. B) and E) for $Re=90, 000$. C) and F) for $Re=120, 000$.}
    \label{fig:fig11}
\end{figure}

Figure \ref{fig:fig11} presents the results obtained during the closed-loop control implementation at three different incoming velocities. Dataset at each $Re$ was recorded three times, and the instantaneous $C_D$ obtained as the average of the three trials (black) is displayed, together with the uncertainty ($\delta C_D$), obtained as standard error ($\sigma/\sqrt{N}$, where $\sigma$ is the standard deviation and $N$ is the number of repetitions). To provide a better interpretation of the effects of deploying dimples on the instantaneous forces, we also introduce a moving average (red line) calculated with a window size of $1s$ (1000 samples). As explained before, we start the procedure by acquiring data at $k/d=0.000$ for one minute. Then, the incoming flow velocity is estimated by using the instantaneous force data and assuming $C_D=0.55$. We observe a good agreement between the input $Re$ ($60,000$, $90, 000$ and $120, 000$) and the estimated conditions, with the largest discrepancy noted at the lowest velocity, for an overall error of less than 10\%. Once the $Re$ is estimated, the control approach evaluates the optimum dimple depth for drag reduction with Equation~\ref{Eq4}, leading to a $k/d_{opt}$ of $0.0075$, $0.0035$ and $0.0022$ for each of the $Re$, in line with the forces evolution presented at Figure \ref{fig:fig6}. Once the dimples are deployed (the green shadow region in the figure denotes the transition period), a large $C_D$ reduction of 40\% is achieved for all incoming velocities. The real-time application capabilities of our smart morphable sphere is underscored when considering that the overall transition time ranges from $3.5s$ to $0.8s$ at the highest $Re$, leading to an almost instant performance benefit. Although this closed-loop control is only valid for fixed velocities, it serves as a proof-of-concept demonstration of the advantages of a smart morphing surface when compared to spheres with rigid dimples, and paves the way to more generic implementations, able to deal with unsteady incoming conditions, such as gusts, wakes, or varying flow velocities. 

\section{Summary and Conclusions}\label{sec4}




In this study, we developed a smart morphable sphere that allows precise control of the surface topography. A comprehensive series of systematic experiments were conducted for Reynolds numbers varying in the range $5\times10^4 \leq Re \leq 1.3\times10^5$ and dimple depth ratios of $0 \leq k/d \leq 2\times10^{-2}$ employing simultaneous force and flow field measurements.\\


It was found that the dimple depth ratio plays a key role in affecting the onset of drag crisis as well as the minimum drag that could be achieved. While the baseline case ($k/d=0.000$) resulted in a constant $C_D\approx0.5$, introducing dimples led to $C_D$ reductions of up to 40\% across all $Re$  considered. We observed that shallower dimples were more effective in reducing drag at high $Re$, with $k/d=0.002$ triggering a drag crisis at $Re=130,000$, resulting in a drastic 50\% reduction in $C_D$ compared to a smooth surface. As $k/d$ increased, the critical $Re$ value shifted to lower Reynolds numbers of $Re=70,000$ at $k/d=0.006$, while drag reduction was capped at 40\%. Further increases in $k/d$ led to a reversal in the drag reduction trend, with performance benefits for $k/d=0.02$ limited to 9\%.\\

2D-2C Particle image velocimetry was conducted to correlate force evolution with wake characteristics, revealing that drag reductions were associated with a delay in the global flow separation location over the sphere's surface. For instance, $k/d=0.004$ exhibited a separation angle of $\theta_s=147^\circ$ at $Re=90,000$, contrasting with $\theta_s=97^\circ$ for $k/d=0.000$ under the same conditions. Finally, the experimental data collected in this study facilitated the development of a model linking $Re$ to optimal $k/d$ for minimizing drag across the entire range of Reynolds number tested. This model delineates the capability of the morphable approach on minimizing $C_D$ by up to 50\% across the entire range of conditions examined. To prove the real-time implementation capabilities of the smart morphable sphere, we conducted a first closed-loop implementation able to deploy optimum dimple depths for a given incoming flow condition, showing an almost instant drastic $C_D$ reduction of 40\% at three $Re$ values.\\

This study underscores the significance of properly choosing $k/d$ to minimize drag for a given $Re$, as any departure from the optimal dimple depth can quickly diminish or nullify performance gains. Nonetheless, further research could supplement some of the study's findings, such as the reversal in drag reduction trends as $k/d$ increases for a fixed $Re$ value. Additionally, experimental limitations have restricted the $Re$ range of this study, but future studies could extend the range of incoming conditions for which our smart morphable sphere can offer drag benefits.\\

This study not only advances our understanding of how roughness parameters influence drag reduction around a sphere but also establishes the groundwork for implementing adaptive flow control strategies to achieve drag reduction across various Reynolds numbers. These strategies can be directly applied to small marine and aerial unmanned vehicles. Further exploration into alternative morphing mechanisms, such as shape memory polymers, is encouraged for future research. \\


\textbf{Acknowledgements}\\

The authors would like to acknowledge the Department of Aerospace Engineering at the University of Michigan for kindly allowing us to use their wind tunnel. Additionally, we would like to acknowledge the undergraduate student, Aditya Madhukar, who helped develop the preliminary pneumatic actuation system for the morphing mechanism at the initial stages.  \\

\textbf{Declaration of Interests.}\\

The authors report no conflict of interest.

\bibliography{dimplepapers}

\end{document}